\documentstyle[psfig]{mn}

\input{epsf}

\def \bj {b$_{\rm J}$}

\def \sqdeg {$\Box^{\circ}$}

\def \etal {et al.}

\def \bj {\rm b$_{\rm J}$}

\def \date {March 2001}

\def\lsim{\mathrel{\hbox{\rlap{\hbox{\lower4pt\hbox{$\sim$}}}\hbox{$<$}}}}
\def\gsim{\mathrel{\hbox{\rlap{\hbox{\lower4pt\hbox{$\sim$}}}\hbox{$>$}}}}

\def\and   {\rm {et al.} \rm}  % Roman Font is the MNRAS/ApJ style these days
\def\etal  {\rm {et al.} \rm}

\begin{document}

%TITLES AND AUTHORS
\title[The 2dF Galaxy Redshift Survey: Luminosity dependence of galaxy clustering]
{
\vspace{-0.75cm}The 2dF Galaxy Redshift Survey: Luminosity dependence of galaxy clustering}
\vspace{-0.5cm}

\author[Norberg et al.]{
\parbox[t]{\textwidth}{
\vspace{-0.75cm}
Peder Norberg$^{1}$,
Carlton M.\ Baugh$^{1}$,
Ed Hawkins$^{2}$,
Steve Maddox$^{2}$,
John A.\ Peacock$^{3}$,
Shaun Cole$^{1}$, 
Carlos S.\ Frenk$^{1}$, 
Joss Bland-Hawthorn$^{4}$,
Terry Bridges$^{4}$, 
Russell Cannon$^{4}$, 
Matthew Colless$^{5}$, 
Chris Collins$^{6}$, 
Warrick Couch$^{7}$, 
Gavin Dalton$^{8}$,
Roberto De Propris$^{7}$,
Simon P.\ Driver$^{9}$, 
George Efstathiou$^{10}$, 
Richard S.\ Ellis$^{11}$, 
Karl Glazebrook$^{12}$, 
Carole Jackson$^{5}$,
Ofer Lahav$^{10}$, 
Ian Lewis$^{4}$, 
Stuart Lumsden$^{13}$, 
Darren Madgwick$^{10}$,
Bruce A.\ Peterson$^{5}$, 
Will Sutherland$^{3}$,
Keith Taylor$^{4,11}$
}
\vspace*{6pt} \\ 
$^{1}$Department of Physics, University of Durham, South Road, 
    Durham DH1 3LE, UK \\ 
$^{2}$School of Physics \& Astronomy, University of Nottingham,
       Nottingham NG7 2RD, UK \\
$^{3}$Institute for Astronomy, University of Edinburgh, Royal Observatory, 
       Blackford Hill, Edinburgh EH9 3HJ, UK \\
$^{4}$Anglo-Australian Observatory, P.O.\ Box 296, Epping, NSW 2121,
    Australia\\  
$^{5}$Research School of Astronomy \& Astrophysics, The Australian 
    National University, Weston Creek, ACT 2611, Australia \\
$^{6}$Astrophysics Research Institute, Liverpool John Moores University,  
    Twelve Quays House, Birkenhead, L14 1LD, UK \\
$^{7}$Department of Astrophysics, University of New South Wales, Sydney, 
    NSW 2052, Australia \\
$^{8}$Department of Physics, University of Oxford, Keble Road, 
    Oxford OX1 3RH, UK \\
$^{9}$School of Physics and Astronomy, University of St Andrews, 
    North Haugh, St Andrews, Fife, KY6 9SS, UK \\
$^{10}$Institute of Astronomy, University of Cambridge, Madingley Road,
    Cambridge CB3 0HA, UK \\
$^{11}$Department of Astronomy, California Institute of Technology, 
    Pasadena, CA 91125, USA \\
$^{12}$Department of Physics \& Astronomy, Johns Hopkins University,
       Baltimore, MD 21218-2686, USA \\
$^{13}$Department of Physics, University of Leeds, Woodhouse Lane,
       Leeds, LS2 9JT, UK \\
\vspace*{-1.cm}}
\maketitle 
 
\begin{abstract}
We investigate the dependence of the strength of galaxy clustering on 
intrinsic luminosity using the Anglo-Australian two degree field 
galaxy redshift survey (2dFGRS). The 2dFGRS is over an order of 
magnitude larger than previous redshift surveys used to address 
this issue. We measure the projected two-point correlation function of 
galaxies in a series of volume-limited samples. 
The projected correlation function is free from any distortion 
of the clustering pattern induced by peculiar motions and is 
well described by a power-law in pair separation over the range 
$ 0.1 < (r/\,h^{-1}\,{\rm Mpc}) < 10 $. 
The clustering of $L^{*}$ ($M_{b_{J}}-5\log_{10}h = -19.7$) galaxies 
in real space is well fit by a correlation length  
$r_{0} = 4.9 \pm 0.3 \, h^{-1}\,{\rm Mpc}$ and power-law slope  
$\gamma = 1.71 \pm 0.06$. 
The clustering amplitude increases slowly with absolute magnitude  
for galaxies fainter than $M^{*}$, 
but rises more strongly at higher luminosities. 
At low luminosities, our results agree with measurements from the SSRS2 
by Benoist \etal However, we find a weaker dependence of clustering 
strength on luminosity at the highest luminosities. 
The correlation function amplitude increases by a factor of 4.0 
between $M_{b_{J}} -5\log_{10}h = -18$ and $-22.5$, and the
most luminous galaxies are 3.0 times more strongly clustered
than $L^*$ galaxies.
The power-law slope of the correlation function shows remarkably 
little variation for samples spanning a factor of 20 in luminosity.
Our measurements are in very good agreement with the predictions 
of the hierarchical galaxy formation models of Benson et al. 
\end{abstract}


\section{Introduction} 
\label{s:intro}

A major obstacle to be overcome by any successful theory of the formation 
of large scale structure is the problem of how galaxies trace 
the distribution of matter in the Universe.
Measurements of differential galaxy clustering as a function of 
colour (Willmer \etal 1998), morphological type (Davis \& Geller 1976; 
Iovino \etal 1993) and selection passband (Peacock 1997; Hoyle \etal 1999) 
imply the existence of biases between the distribution 
of galaxies and of mass. 

A generic prediction of hierarchical structure formation models is that 
rarer objects should be more strongly clustered than average 
(Davis \etal 1985; White \etal 1987). Correspondingly, if more luminous 
galaxies are associated with more massive haloes, then these galaxies are  
expected to exhibit stronger clustering than the galaxy 
population as a whole (for the special case of bright galaxies 
at high redshift, see for example Baugh \etal 1998; Governato \etal 1998).
However, the form of the dependence of the amplitude 
of galaxy clustering on luminosity remains controversial even after 
more than twenty years of constructing and analysing redshift surveys 
of the local Universe. 
In the literature, claims of a dependence of galaxy 
clustering on luminosity (e.g. Davis \etal 1988; Hamilton 1988; 
Maurogordato \& Lachieze-Rey 1991; Park \etal 1994; Benoist \etal 1996; Willmer 
\etal 1998; Guzzo \etal 2000) have been made with 
similar regularity to claims of non-detections (e.g. Phillips \& Shanks 1987;  
Hasegawa \& Umemura 1993; Loveday \etal 1995; Szapudi \etal 2000; 
Hawkins \etal 2001).
Part of the reason for this disagreement is a mismatch 
in the range of luminosities and clustering length scales 
considered in earlier studies.
However, the main problem with earlier work is the small size of the 
redshift surveys analysed, both in terms of volume and number of galaxies.
With previous surveys, the dynamic range in luminosity for which 
clustering can be measured reliably is limited, particularly when 
volume-limited samples are used. 
Due to the small volumes probed, it has generally not been possible to 
compare the clustering of galaxies of different luminosity measured 
within the same volume. 
These results have generally been affected by sampling fluctuations 
that are difficult to quantify.
This problem is compounded by underestimation of the errors 
on the measured correlation functions and on the power-law fits 
traditionally employed in this subject. 

In this paper, we use the largest extant local survey, the 
Anglo-Australian two degree field galaxy redshift survey 
(hereafter 2dFGRS), to address the issue of how clustering depends 
upon galaxy luminosity. We describe the 2dFGRS  and the construction of 
volume-limited samples in Section \ref{s:data}, and our estimation 
of the correlation function is described in Section \ref{s:meth}. 
Our results for the real space correlation function are given in Section 
\ref{s:res}. We compare our results with those from previous studies 
and with the predictions of simulations of hierarchical galaxy 
formation in Section \ref{s:end}.

\section{The data}
\label{s:data}

\subsection{The 2dFGRS sample}

The 2dFGRS is selected in 
the photometric \bj\ band from the APM galaxy survey 
(Maddox \etal 1990a,b; 1996) and its subsequent extensions  
(Maddox et al. in preparation). 
The survey is divided into two regions and covers approximately 2000 
square degrees. 
The bulk of the solid angle of the survey is made up of two 
broad strips, one in the South Galactic Pole region (SGP) covering  
approximatively $-37^\circ\negthinspace.5<\delta<-22^\circ\negthinspace.5$,  
$21^{\rm h}40^{\rm m}<\alpha<3^{\rm h}30^{\rm m}$ and the other in the 
direction of the North Galactic Pole (NGP), spanning  
$-7^\circ\negthinspace.5<\delta<2^\circ\negthinspace.5$,  
$9^{\rm h}50^{\rm m}<\alpha<14^{\rm h}50^{\rm m}$. 
In addition to these contiguous regions, there are 
a number of circular 2-degree fields 
scattered pseudo-randomly over the full extent of the low extinction regions 
of the southern APM galaxy survey. 
In this paper, we use the redshifts obtained prior to 
January 2001, over $160\,000$ in total. 
As we are mainly interested in measuring clustering out to separations 
of order $20 \,h^{-1}\,$Mpc, we do not include galaxies that lie in the 
random fields in our analysis.

In order to select an optimal sample for the measurement of  
the two point correlation function, we apply a weighting scheme to 
objects in the 2dFGRS. 
A weight is assigned to each measured redshift based upon the redshift 
completeness mask, the construction of 
which is explained in Colless \etal (2001; see also Norberg \etal 2001, 
in preparation). 
We require a relatively high completeness in a given direction on 
the sky, so that, in practice, 
our results are fairly insensitive to the precise details of the 
weighting scheme. 
Excluding areas below our completeness threshold (which arise mainly 
as a result of the tiling strategy adopted to make optimal use 
of telescope time, 
coupled with the fact that the survey is not yet finished), we 
estimate the effective solid angle used in the 
SGP region is $\sim$ 420 \sqdeg, 
and in the NGP $\sim$ 190 \sqdeg.

\subsection{Constructing a volume-limited sample}

\begin{table*}
\caption{Properties of the combined NGP \& SGP volume-limited sub-samples 
analysed. The second column gives the median magnitude of each sample. Columns 6 and 7 
list the best fitting correlation length, $r_{0}$, 
and power-law slope $\gamma$ of the correlation function in real space, fitted 
over the range $0.5 \le \sigma/(\,h^{-1}\,{\rm Mpc}) \le 10$. Column 8 gives the 
value of $A(\gamma)$, defined by eqn \ref{eq:pow}, evaluated for the 
best fitting value of $\gamma$.}
\begin{tabular}{cccccccc}   
\hline
Mag. range           & Median magnitude   &  $N_{gal}$ & $z_{\rm min}$   & $z_{\rm max}$ & $r_{0}$         & $\gamma$ & 
$A(\gamma)$ \\
$M_{b_{J}}-5\log_{10} h$  & $M_{b_{J}}-5\log_{10} h$       &            &                 &               & ($\,h^{-1}\,$Mpc)  &     &    \\ \hline
$-18.0 \quad  -18.5$ & $-18.11$            &  \phantom{0}7061      &  0.010          &   0.086       & $4.14 \pm 0.64$ & $1.78 \pm 0.10$  & 3.75  \\
$-18.5 \quad  -19.0$ & $-18.61$            &  \phantom{0}9382      &  0.013          &   0.104       & $4.43 \pm 0.45$ & $1.75 \pm 0.08$   &3.80   \\
$-19.0 \quad  -19.5$ & $-19.11$            & 13690      &  0.016          &   0.126       & $4.75 \pm 0.44$ & $1.68 \pm 0.08$  &  4.14 \\
$-19.5 \quad  -20.0$ & $-19.60$            & 15123      &  0.020          &   0.152       & $4.92 \pm 0.27$ & $1.71 \pm 0.06$  &  4.01 \\
$-20.0 \quad  -20.5$ & $-20.09$            & 13029      &  0.025          &   0.182       & $5.46 \pm 0.28$ & $1.68 \pm 0.06$  &  4.14  \\
$-20.5 \quad  -21.0$ & $-20.58$            &  \phantom{0}9114      &  0.031          &   0.220       & $6.49 \pm 0.29$ & $1.63 \pm 0.06$   &4.39  \\
$-21.0 \quad  -21.5$ & $-21.06$            &  \phantom{0}3644      &  0.039          &   0.270       & $7.58 \pm 0.48$ & $1.76 \pm 0.09$   &3.82   \\\hline

$-18.0 \quad  -19.0$ & $-18.22$  &  12594      &  0.013          &   0.086       & $4.06 \pm 0.53$ & $1.79 \pm 0.09$  &  3.72 \\
$-19.0 \quad  -20.0$ & $-19.19$ &  21874      &  0.020          &   0.126       & $4.75 \pm 0.44$ & $1.70 \pm 0.08$  &  4.06 \\
$-20.0 \quad  -21.0$ & $-20.13$ &  17383      &  0.031         &   0.182       & $5.65 \pm 0.30$ & $1.69 \pm 0.06$   &  4.10\\
$-21.0 \quad  -22.0$ & $-21.07$ &  \phantom{0}4013      &  0.048          &   0.270       & $8.12 \pm 0.46$ & $1.78 \pm 0.12$   &3.75 \\
$-21.5 \quad  -22.5$ & $-21.55$ &  \phantom{0}1002      &  0.059          &   0.280       & $9.38 \pm 1.48$ & $1.69 \pm 0.15$   &4.10  \\ \hline
\end{tabular}
\end{table*}

In this paper, we analyse a series of volume-limited 
subsamples drawn from the 2dFGRS. The advantage of this 
approach is that the radial selection function is uniform, 
and the only variations in the space density of galaxies within   
each volume are due to clustering. 
By contrast, in a flux-limited survey, the galaxy number 
density is a strong function of radial distance and this  
needs to be corrected for when measuring the clustering. 
The disadvantage of using a volume-limited sample is that a large 
number of galaxies in the flux-limited survey do {\it not\/} 
satisfy the selection cuts (which are explained below).
This is a serious problem for previous surveys, but not for 
a survey the size of the 2dFGRS. 
As we demonstrate in Section \ref{s:res}, the volume-limited samples 
we analyse give robust clustering measurements and contain over an order 
of magnitude more galaxies than similar samples constructed from previous 
surveys (see Table 1). 

The construction of a volume-limited sample drawn from a flux-limited 
redshift survey requires a range of absolute magnitudes 
to be specified. Since a  
flux-limited survey has both bright and faint apparent magnitude 
limits, the selected range of absolute magnitudes requires that both 
a minimum ($z_{\rm min}$) and a maximum ($z_{\rm max}$) redshift 
cut be applied to the volume-limited sample. 
Thus, in principle, a galaxy included in the volume-limited 
sample could be displaced to any redshift between 
$z_{\rm min}$ and $z_{\rm max}$ and still remain within the 
bright and faint {\it apparent\/} magnitude limits of the flux-limited 
survey. 

In order to estimate the absolute magnitude of 2dFGRS galaxies 
at redshift zero, it is necessary to apply corrections for band 
shifting ($k-$correction) and evolution in the stellar 
populations ($e-$correction). 
We adopt a global $k+e$ correction given by 
$k + e = 0.03 z/\left(0.01 + z^{4}\right)$, which is a 
good fit to the correction calculated for the \bj\ selected ESO 
Slice Project survey using population synthesis models (see Fig. 1 of 
Zucca \etal 1997). 
This form for the $k+e$ correction gives consistent luminosity 
functions for the 2dFGRS when the survey is divided into redshift bins, 
indicating that it adequately accounts for the degree of 
evolution in galaxy luminosity over the lookback time spanned 
by the survey (Norberg \etal 2001, in preparation).
Our results are unchanged if we use the mean of the $k$ corrections 
for different spectral types given by Madgwick \etal (2001, in preparation).
The values of $z_{\rm min}$ and $z_{\rm max}$ that define a volume-limited 
sample drawn from the 2dFGRS vary slightly with position 
on the sky. This is due to revisions made to the map of galactic 
extinction (Schlegel, Finkbeiner \& Davis 1998) and to the CCD calibration 
of APM plate zero points since the definition of the original input 
catalogue.
Throughout the paper, we adopt an $\Omega_{0}=0.3$, $\Lambda_{0}=0.7$ 
cosmology to convert redshift into comoving distance. 

\section{Estimating the two-point correlation function}
\label{s:meth}
\begin{figure}
{\epsfxsize=8.2truecm \epsfysize=11.6truecm 
\epsfbox[58 188 464 678]{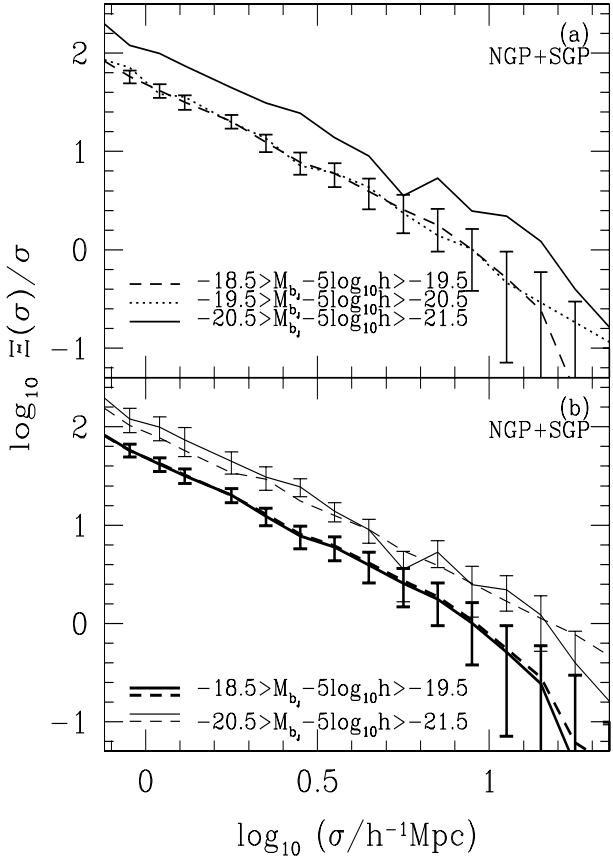}}
\caption{
(a) The projected correlation function measured for galaxies 
in three different absolute magnitude bins in the {\it same\/} volume. 
The faintest sample contains $16\,134$ galaxies, the middle sample contains 
$6\,186$ galaxies and the brightest sample contains $985$ galaxies. 
For clarity, error bars are plotted only on the correlation function 
of galaxies with $-18.5 \ge M_{b_{J}}-5\log_{10} h \ge -19.5$.
(b) A comparison of the correlation function of galaxies in the 
same absolute magnitude bins but measured in different (although not 
completely independent) volumes.
The heavy lines show results for galaxies with 
$-18.5 \ge M_{b_{J}}-5\log_{10} h \ge -19.5$ and the light lines show 
results for a brighter bin with $-20.5 \ge M_{b_{J}}-5\log_{10} h \ge -21.5$. 
In each case, the dashed line shows the estimate from the optimal sample 
(see text) for the selected magnitude bin, 
whilst the solid line shows an estimate 
of the correlation function from the volume analysed in 
Fig. \ref{fig1}(a). For the $-20.5 \ge M_{b_{J}}-5\log_{10} h \ge -21.5$ 
magnitude bin, the optimal estimate is measured using $10\,962$ galaxies, 
which should be contrasted with the $985$ galaxies used to make the 
measurement shown by the light solid line, in a volume 
defined by a broader magnitude bin.
}
\label{fig1}
\end{figure}

The galaxy correlation function is estimated on a two dimensional 
grid of pair separations parallel ($\pi$) and perpendicular ($\sigma$) 
to the line-of-sight. 
To estimate the mean density of pairs, a catalogue of unclustered points 
is generated with the same angular selection and  
($z_{\rm min}$, $z_{\rm max}$) values as the data. 
The correlation function is estimated by 
\begin{equation}
\xi = \frac{ DD - 2DR + RR }{RR}, 
\end{equation}
where $DD$, $DR$ and $RR$ are the suitably normalised number of weighted 
data-data, data-random and random-random pairs respectively in 
each bin (Landy \& Szalay 1993).

Contours of constant clustering amplitude in the redshift space correlation 
function, $\xi(\sigma,\pi)$, are distorted as a result of the peculiar motions 
of galaxies, as demonstrated for the 2dFGRS by Peacock \etal (2001). 
On small scales, random motions inside virialised structures 
elongate the constant-$\xi$ contours in the $\pi$ direction, whereas on large 
scales, coherent flows flatten the contours. The latter effect 
was measured clearly for the first time for galaxies using the 2dFGRS 
(Peacock \etal 2001).  
The dependence of the redshift space correlation function on galaxy 
luminosity is analysed in a separate paper (Hawkins \etal 2001, 
in preparation).
In this paper, to simplify the interpretation, 
we consider only clustering in real space, which we infer by projecting 
the measured correlation function along the line-of-sight. 
We compute a dimensionless quantity, $\Xi(\sigma)/\sigma$ by integrating over 
the measured $\xi(\sigma,\pi)$ grid (note that $\Xi(\sigma)$ is sometimes 
referred to as $w(r_{p})$ in the literature): 

\begin{equation} 
\frac{\Xi(\sigma)}{\sigma} = \frac{1}{\sigma} \int_{-\infty}^{\infty} \xi(\sigma,\pi) {\rm d}\pi. 
\end{equation}
In practice, the integral converges by a pair separation 
of $\pi = 75 \,h^{-1}\,$Mpc.
The projected correlation function can, in turn, be written as an integral 
over the spherically averaged real space correlation function, $\xi(r)$, 
\begin{equation} 
\frac{\Xi(\sigma)}{\sigma} = \frac{2}{\sigma} \int_{\sigma}^{\infty} \xi(r) 
\frac{r{\rm d}r}{\left(r^{2}-\sigma^{2}\right)^{1/2}}, 
\end{equation}
(Davis \& Peebles 1983). 
If the real space correlation function is a 
power-law (which is a reasonable approximation for APM galaxies 
out to separations around $r \sim 10 \,h^{-1}\,$Mpc, see e.g. Baugh 1996), then 
\begin{equation} 
\frac{\Xi(\sigma)}{\sigma} = \left(\frac{r_{0}}{\sigma} \right)^{\gamma}
\frac{\Gamma(1/2)\Gamma( [\gamma-1]/2)}
{\Gamma(\gamma/2)} = \left(\frac{r_{0}}{\sigma} \right)^{\gamma} A(\gamma), 
\label{eq:pow}
\end{equation}
where $\xi(r)=\left(r_{0}/r\right)^{\gamma}$ and $r_{0}$ is the correlation length.

Previous studies have estimated the error on the measured correlation function 
from the Poisson statistics of the pair counts in each bin (Peebles 1980) or 
by bootstrap resampling of the data (e.g. Benoist \etal 1996). Since we study 
a range of samples corresponding to different luminosity bins  and also 
compare samples from different volumes, it is important to include an estimate 
of the sampling fluctuations in the error budget for the correlation function. 
This we derive from  analysis of 22 mock 2dFGRS catalogues 
constructed from the $\Lambda$CDM Hubble Volume dark matter simulation, 
in the manner explained by Baugh \etal (2001, in preparation; 
see also Cole \etal 1998). 
In order to mimic the clustering of the 2dFGRS, a biasing scheme 
is employed to select particles in the simulations with a probability which 
is a function of  the final dark matter density field, smoothed 
with a Gaussian filter (model 2 of Cole \etal 1998).
The mock catalogues have the same clustering amplitude as galaxies 
in the flux-limited 2dFGRS and the same selection criteria that are 
applied to the data are used in the construction of the mock surveys. 
The clustering amplitude in the mocks is independent of luminosity. 
The error bars that we plot on correlation functions measured from the 2dFGRS 
are the {\it rms\/} found by averaging over the 22 mock catalogues.

\begin{figure}
{\epsfxsize=7.9truecm \epsfysize=7.9truecm 
\epsfbox[40 210 540 627]{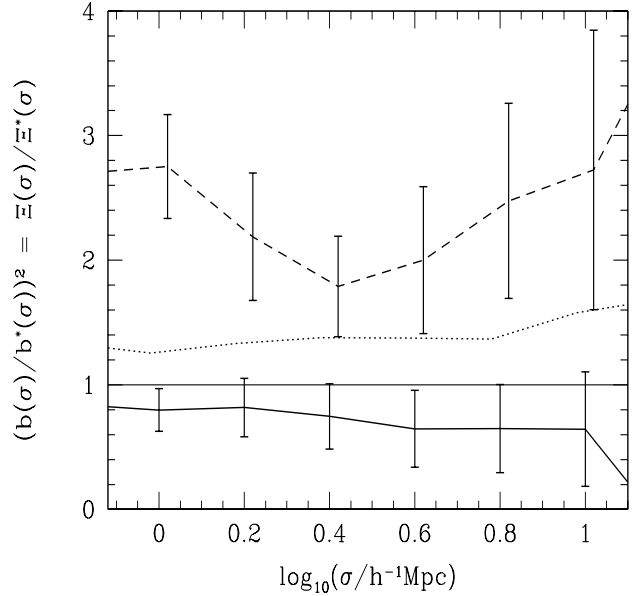}}
\caption{
The ratio of the projected correlation function of 
galaxies in different magnitude slices to the projected 
correlation function of galaxies with 
$-19 \ge M_{b_{J}}-5\log_{10} h \ge -20$. 
Note that the ratio is plotted on a linear scale, whilst the pair 
separation is on a log scale.
The solid line shows the ratio for galaxies with absolute 
magnitudes in the range $-18 \ge M_{b_{J}}-5\log_{10} h \ge -19$, 
the dotted line for $-20 \ge M_{b_{J}}-5\log_{10} h \ge -21$ and the 
dashed line for $-21 \ge M_{b_{J}}-5\log_{10} h \ge -22$.
For clarity, error bars have been omitted from the dotted line but 
these are comparable in size with those plotted on the solid line.
}
\label{fig2}
\end{figure}

\section{Results}
\label{s:res}

\begin{figure*}
{\epsfxsize=17.truecm \epsfysize=14.truecm 
\epsfbox[0 210 568 672]{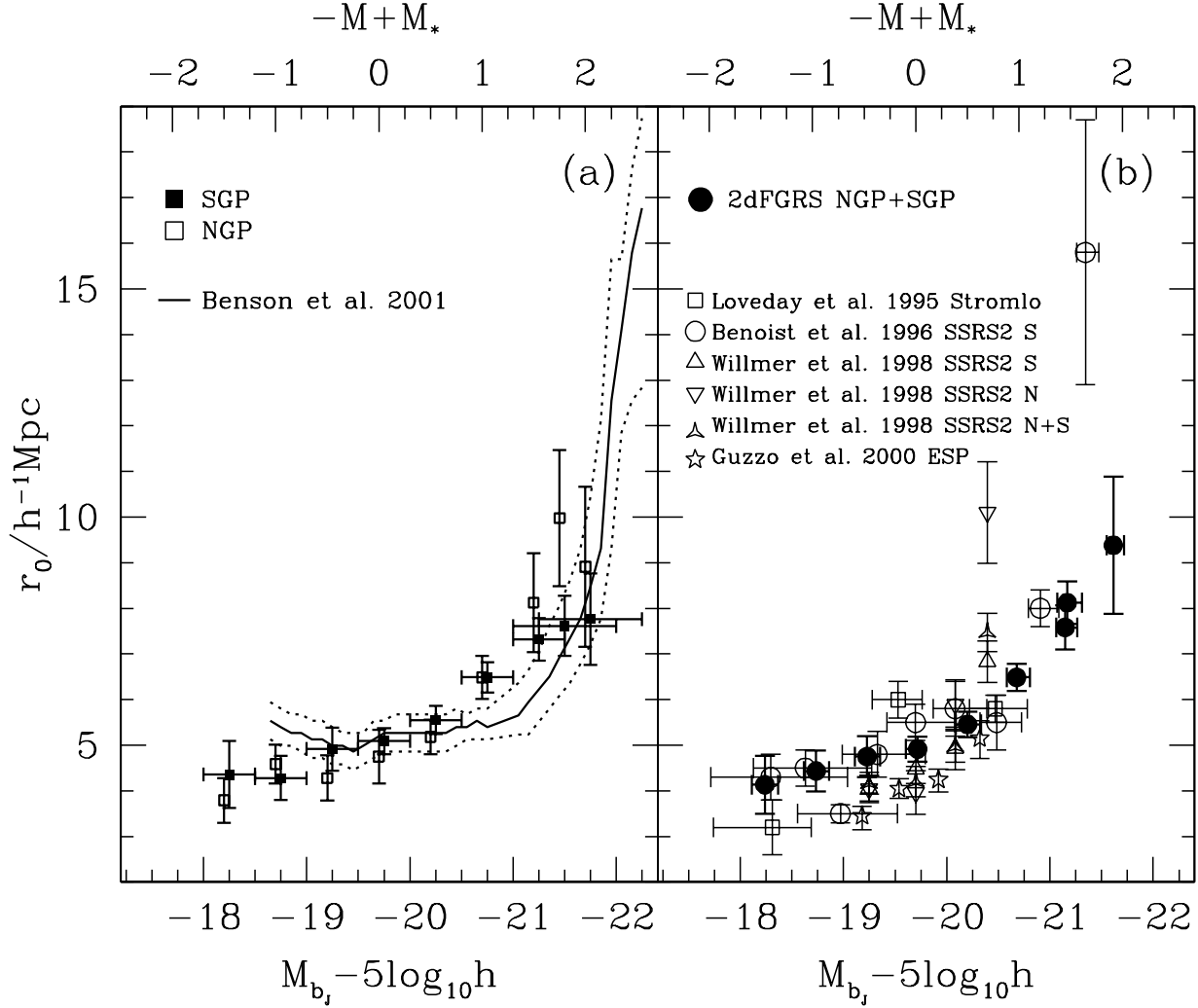}}
\caption{
(a) The correlation length in real space as a function of 
absolute magnitude. Results are shown for the 
SGP and NGP regions separately.  
The NGP points are plotted with an offset of 0.05 mag for 
clarity. 
Horizontal error bars on the SGP points indicate 
the absolute magnitude range of each bin, and each point is 
plotted at the bin centre. 
In both cases, the brightest data points are for galaxies in 
one magnitude wide bins. 
The solid line shows the predictions of the semi-analytic 
model of Benson \etal (2001), computed in a series of overlapping 
bins, each $0.5$ magnitudes wide. The dotted curves show an 
estimate of the errors on this prediction, including the sample variance 
expected for a volume equal to that of the N-body simulation used.
(b) The real space correlation length estimated combining pairs counts 
in the NGP and SGP (filled circles). 
The open symbols show a selection of recent data from other studies.
The data for surveys selected in the $B$-band have been 
corrected to the \bj\ band using the approximate 
relation $M_{b_{J}} = M_{B} - 0.2$.
In order to compare samples defined by cumulative and differential 
magnitude bins, the data points 
are plotted at the median magnitude of each sample.
}
\label{fig3}
\end{figure*}

\begin{figure}
{\epsfxsize=8.2truecm \epsfysize=8.2truecm 
\epsfbox[25 200 546 710]{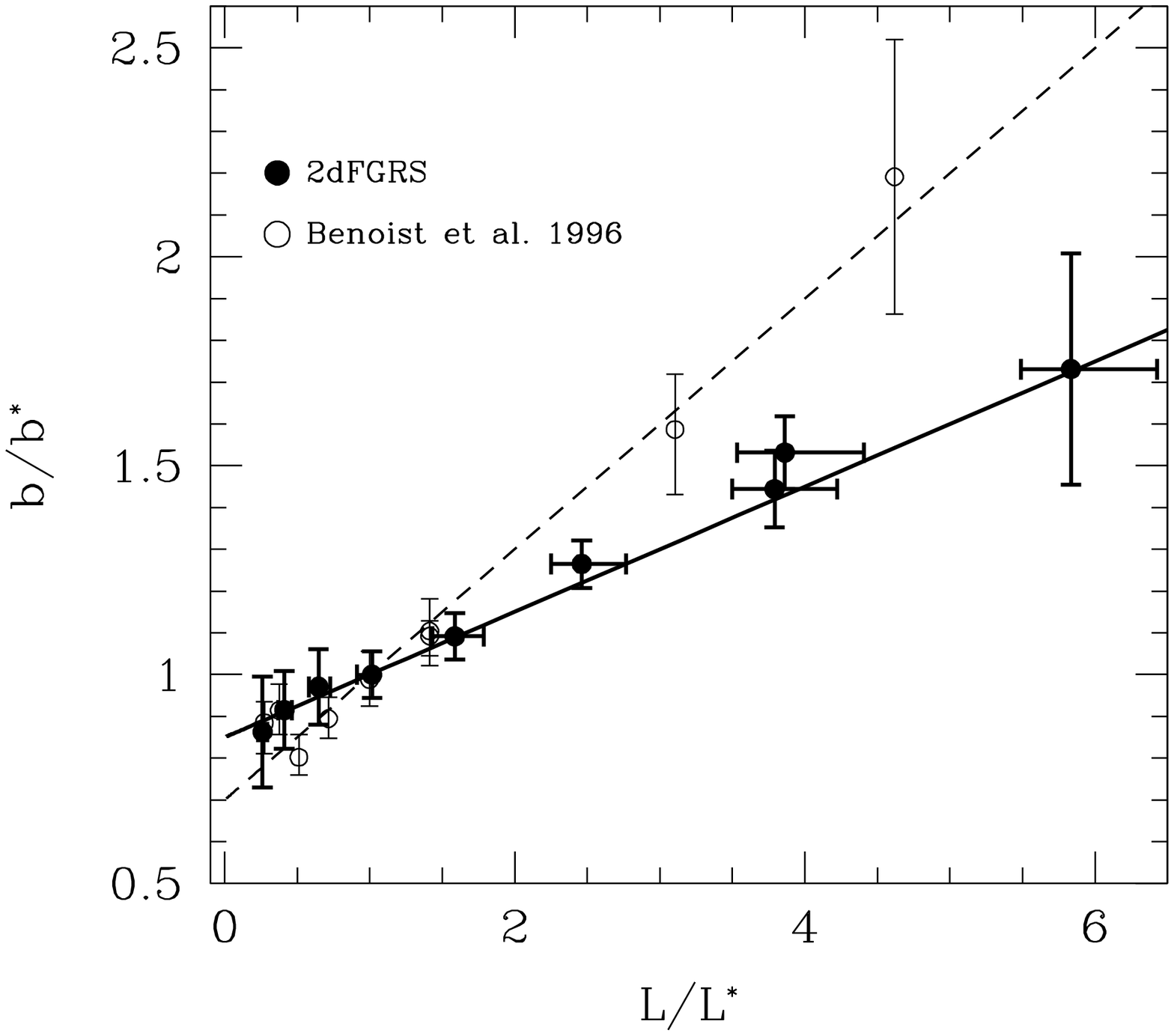}}
\caption{
The variation of the relative bias as a function of luminosity, 
using the clustering of $L^{*}$ galaxies as a reference point (see text 
for definition).
The 2dFGRS points are plotted at the median magnitude of each sample 
and the horizontal bars show the quartile magnitude range.
The Benoist \etal (1996) points are taken from their Fig. 5 and are 
plotted at the median value of $L/L^{*}$ for each sample. 
Note that the error bars on the Benoist \etal points are obtained 
by averaging over correlated bins in pair separation.
The curves shows parametric fits: the Benoist \etal  measurements 
are well fitted by $b/b^{*} = 0.7 + 0.3 L/L^{*}$ (dashed line), whereas 
the 2dFGRS results suggest a more modest dependence on luminosity: 
$b/b^{*} = 0.85 + 0.15 L/L^{*}$ (solid line). 
}
\label{fig4}
\end{figure}

We first demonstrate the robustness of the approach of measuring the 
correlation function in volume-limited samples. 
Unless stated otherwise, we have added the pair counts in the NGP and SGP 
regions to compute correlation functions. 
In Fig. \ref{fig1}(a), we show the correlation function of galaxies 
in three disjoint absolute magnitude bins measured in the same volume. 
The sampling fluctuations are 
therefore virtually the same for each sub-sample, although the number of galaxies varies 
between them. 
There is a clear difference in the clustering amplitude of  
galaxies in the brightest absolute magnitude bin.  
Next, we demonstrate that sampling fluctuations are not important in a 
survey the size of the 2dFGRS. 
We show, in Fig. \ref{fig1}(b), the 
correlation function in two fixed absolute magnitude bins measured in  
different volume-limited sub-samples.
Specifically, the dashed lines show the correlation function for  
the optimal volume-limited sample, appropriate to the selected absolute 
magnitude bin. 
Such a sample contains the maximum number of galaxies 
in that magnitude bin.
The different estimates of the correlation function 
agree within the errors.

We now focus attention on the series of volume-limited 
subsamples covering the range $-18 \ge M_{b_{J}}-5 \log_{10} h \ge -22.5$, 
whose characteristics are listed in Table 1.  
The shape and amplitude of the projected correlation function 
in a selection of these samples is compared in Fig. \ref{fig2} 
with the correlation function of galaxies in the magnitude range 
$-19 \ge M_{b_{J}}-5\log_{10} h \ge -20$.
The shape of the correlation function varies relatively little 
with the absolute magnitude that defines the sample in contrast to 
the amplitude of the correlation function, which changes significantly 
for the brightest magnitude slice.  
Another view of this trend is given in Fig. \ref{fig3}(a) where  
we plot the real space correlation length as a function of 
absolute magnitude. 
The best fitting values of the correlation length, $r_{0}$, and 
power-law slope $\gamma$, are determined by applying eqn. \ref{eq:pow} 
to the measured correlation function over the pair separation range 
$0.5 \le \sigma/(\,h^{-1} \,{\rm Mpc}) \le 10$ and carrying out a 
$\chi^{2}$ minimisation. 
This simple $\chi^{2}$ approach will not, however, give reliable estimates 
of the errors on the fitted parameters due to the correlation between 
the estimates at differing pair separations. 
We use the mock 2dFGRS catalogues to estimate the errors on the fitted 
parameters. 
In brief, the best fitting values of $r_0$ and $\gamma$ are found for 
each mock individually, using the simple $\chi^{2}$ analysis.
The estimated error bar is the {\it rms}\ scatter in the fitted 
parameters over the ensemble of mock catalogues.

In Fig. \ref{fig3}(a), we plot the correlation lengths 
for the NGP and SGP regions separately. These independent estimates are in 
excellent agreement with one another. The slope of 
the best fitting power-law correlation function, given in Table 1, 
is similar for all the volume-limited samples considered. 
The clustering amplitude increases slowly with luminosity 
for galaxies fainter than $M^{*}$ 
(where $M^{*} = M_{b_{J}}-5\log_{10}h = -19.7$,  
as found by Folkes \etal 1999), 
but rises strongly at higher luminosities. 
The correlation function amplitude increases by a factor of 4.0 
between $M_{b_{J}} -5\log_{10}h = -18$ and $-22.5$, and the
most luminous galaxies are 3.0 times more strongly clustered
than $M^{*}$ galaxies.

\section{Discussion}
\label{s:end}

The volume-limited samples analysed in this paper contain over an 
order of magnitude more galaxies than previous studies of 
the dependence of clustering on galaxy luminosity, allowing 
a more accurate measurement of this effect than was possible before. 
The sheer volume covered by our samples, 
$10^{6} - 2 \times 10^{7}\,h^{-3}\,{\rm Mpc}^{3}$, 
ensures that sampling fluctuations have little impact upon our results. 

We compare the 2dFGRS results with a selection of recent measurements 
taken from the literature since 1995 in Fig. \ref{fig3}(b). 
To compare samples defined by cumulative and differential magnitude bins, 
we plot the datapoints at the median magnitude for the sample, as computed 
using the Schechter function parameters for the 2dFGRS (Folkes \etal 1999).
The horizontal bars plotted on selected points show 
the quartile range of the magnitude distribution in the sample.
Benoist \etal (1996) analysed quasi volume-limited samples in the 
SGP region of the Southern Sky Redshift Survey 2 (SSRS2), 
and found a sharp increase in the correlation length for galaxies brighter 
than $M_{B} - 5\log_{10} h = -20.5$.
The Benoist \etal correlation lengths are measured in redshift space, 
although the authors report that a similar trend with luminosity is seen 
in real space. 
Willmer \etal (1998) re-analysed the SSRS2 South using different volume 
limits and also measured clustering in the SSRS2 North, presenting fits 
for the correlation length in real and redshift space. Intriguingly, 
Willmer \etal find a larger correlation length in real space for 
galaxies with $M_B-5\log_{10} h \sim -20$ than Benoist \etal find in redshift 
space. Moreover, the clear disagreement between the results for the 
brightest galaxies analysed in SSRS2 North and South suggests that 
sampling fluctuations are significant in a survey of this size and 
that the errors on these points have been underestimated (as demonstrated 
in Fig. 4 of Benson \etal 2001). 
Loveday \etal (1995) measured the clustering in real space by 
cross-correlating galaxies in the sparsely-sampled Stromlo/APM redshift survey 
with galaxies in the parent catalogue. Galaxies were considered 
in three absolute magnitude bins. No difference was found between the 
clustering amplitude of $L^{*}$ and super-$L^{*}$ galaxies. 
However, the median magnitude for the most luminous sample considered by 
these authors is only $0.5$ magnitudes brighter than $M^{*}$.
The increase in clustering amplitude with luminosity is 
connected with a change in the mix of morphological types with increasing 
luminosity. The mix of spectral types at the brightest absolute 
magnitudes is dominated by spectra characteristic of elliptical galaxies, 
whereas spiral galaxies are more numerous around $L^{*}$ 
(Folkes \etal 1999; Cole \etal 2001; Madgwick \etal in preparation).
The clustering of galaxies as a function of spectral type will be 
analysed in a separate paper.

Our clustering results can be characterised in a concise way in terms 
of a relative bias parameter, $b/b^{*}$, that gives the amplitude of the 
correlation function relative to that of $L^{*}$ galaxies 
(where $M^{*}=M_{b_{J}}-5\log_{10}h = -19.7$). 
The {\it relative\/} bias between the correlation functions of 
galaxies of different luminosity is assumed to be constant for 
pair separations spanned by the $r_{0}$ values listed in Table 1 
(see also Fig. 2).
The relative bias is then defined by $b/b^{*} = \left(r_{0}/r^{*}_{0}\right)^
{\gamma/2}$, where we take $r^{*}_{0} = 4.9 \pm 0.3 \,h^{-1}\,{\rm Mpc}$ from 
Table 1  and use $\gamma = 1.7$.  
The 2dFGRS results are shown by the filled symbols in Fig. \ref{fig4} 
and are well fitted by the relation $b/b^{*} = 0.85 + 0.15 L/L^{*}$. 
The 2dFGRS data suggest a significantly weaker dependence of the relative 
bias on luminosity than the Benoist \etal data, which follow the relation 
$b/b^{*} = 0.7 + 0.3 L/L^{*}$ (Peacock \etal 2001).
(The parametric fit to the Benoist \etal measurements 
was used by Peacock \etal 2001 to estimate the parameter 
$\beta=\Omega^{0.6}/b$ for $L^{*}$ galaxies in the 2dFGRS. 
Using the above fit to the 2dFGRS measurements changes the inferred 
value for $\beta$ by less than $1 \sigma$ to $\beta = 0.49 \pm 0.08$.)

Hierarchical models of galaxy formation predict that bright 
galaxies should be more strongly clustered than faint galaxies 
(e.g. White \etal 1987; Kauffmann, Nusser \& Steinmetz 1997).
This generic prediction arises because bright galaxies are expected 
to occupy more massive dark matter haloes and these haloes are 
more strongly clustered than the overall distribution of dark matter.
The trend of clustering amplitude with luminosity measured for 
2dFGRS galaxies is in very good agreement with the predictions 
of a simulation of hierarchical galaxy formation taken 
from Fig. 4 of Benson \etal (2001), reproduced as the solid line 
in Fig. \ref{fig3}(a).
In the Benson \etal semi-analytic model, the input parameters are set 
in order to reproduce a subset of local galaxy data, with most emphasis  
given to the field galaxy luminosity function (see Cole \etal 2000). 
No reference is made to clustering data in setting the model parameters.
In a $\Lambda$CDM cosmology, Benson \etal (2000a,b) find excellent agreement 
with the real space correlation function measured for galaxies in 
the APM survey by Baugh (1996). It is remarkable that the same 
model, without any readjustment of parameters, also reproduces the 
dependence of clustering amplitude on luminosity 
exhibited by the  2dFGRS in Fig. \ref{fig3}(a).

\section*{Acknowledgments}
The 2dFGRS is being carried out using the 2 degree field facility on the 
3.9m Anglo\-Australian Telescope (AAT). We thank all those involved in 
the smooth running and continued success of the 2dF and the AAT. 
We thank the referee, Dr. L. Guzzo, for producing a speedy and helpful report, 
and also Andrew Benson for communicating an electronic version of his model 
predictions. 
PN is supported by the Swiss National Science Foundation 
and an ORS award and CMB acknowledges receipt 
of a Royal Society University Research Fellowship. 
This work was supported in part by a PPARC rolling grant at Durham.

\end{document}